\documentclass{article}
\usepackage{color}
\usepackage{subfig}
\usepackage{spconf,amsmath,graphicx,caption,multicol}

\title{End-to-End Multimodal Speech Recognition}
\name{Shruti Palaskar$^{*}$, Ramon Sanabria\sthanks{The first two authors contributed equally to this work.} and Florian Metze}
\address{Carnegie Mellon University; Pittsburgh, PA; U.S.A.\\
\texttt{\{spalaska|ramons|fmetze\}@cs.cmu.edu}}
\begin{document}
\ninept
\maketitle
\begin{abstract}
Transcription or sub-titling of open-domain videos is still a challenging domain for Automatic Speech Recognition (ASR) due to the data's challenging acoustics, variable signal processing and the essentially unrestricted domain of the data. In previous work, we have shown that the visual channel -- specifically object and scene features -- can help to adapt the acoustic model (AM) and language model (LM) of a recognizer, and we are now expanding this work to end-to-end approaches. In the case of a Connectionist Temporal Classification (CTC)-based approach, we retain the separation of AM and LM, while for a sequence-to-sequence (S2S) approach, both information sources are adapted together, in a single model. This paper also analyzes the behavior of CTC and S2S models on noisy video data (How-To corpus), and compares it to results on the clean Wall Street Journal (WSJ) corpus, providing insight into the robustness of both approaches.


\end{abstract}
\section{Introduction}
\label{sec:intro}

Audio-visual speech recognition has been an active area of research for a long time: humans use ``lip-reading'' to improve their robustness against noise, and they control the balance between lip-reading and hearing
adaptively and transparently.

In this paper, we propose to use multi-modal video information slightly differently, and adapt an end-to-end speech recognizer to the visual semantic concepts extracted from a {\em correlated visual scene} that accompanies some speech, for example in a ``How-To'' video. If we see a person standing in a kitchen, holding sliced bread, it is likely that the person is explaining how to make a sandwich, and the acoustic conditions will be comparably clean. If a person is standing in front of a car, it is likely a review of that car, and happening outdoors. Clearly, the language model will be affected as well.

Recognition thus involves two major steps:

\begin{enumerate}
\item For every utterance, extract a visual semantic feature ``context vector'' from a single video frame using deep Convolutional Neural Networks (CNNs) trained for object recognition and scene labeling tasks.

\item Adapt or condition a recognizer to this utterance's context. In this work, we compare a CTC-based system and a S2S system.
\end{enumerate}

This paper builds on earlier work~\cite{miao:is2016,gupta2017visual} and presents first results on multi-modal adaptation of CTC and S2S models. We present various audio-visual adaptation strategies for the CTC model and present our first results with S2S model adaptation. The ultimate goal of this work will be to view automatic speech recognition not primarily as the speech-to-text task, but as a process which sub-titles multi-media material removing repetitions, hesitations or corrections from spontaneous speech as required, much like ``video captioning''~\cite{vinyals2015show}. We show that multi-modal adaptation helps by 2\% absolute improvement in the token error rate. 

While multi-modal adaptation improves recognition in such noisy datasets, we see that there is need for deeper insight into the CTC and S2S models. These models behave very differently with clean, prepared datasets like WSJ than with spontaneous, noisy speech. Ground truth references for the How-To data are less accurate than for WSJ; we see that this influences model training. We present insights into the differences in the output of these two approaches as these issues have not been addressed in prior work yet. Because of the novelty of the S2S approach, we also implement a strong baseline on the WSJ dataset.

In the following sections, we first describe related work in \ref{sec:related}, data and feature extraction in \ref{sec:data}, model descriptions in \ref{sec:models} and results on multimodal adaptation in \ref{ssec:adaptation}, differences in CTC and S2S in \ref{ssec:ctc_s2s} and model differences while using WSJ and How-To in \ref{ssec:wsj_ht}.

\section{Related Work}
\label{sec:related}


On consumer-generated content (like YouTube videos), Deep Neural Network (DNN) models exhibit Word Error Rates (WER) above 40\% \cite{liao2013large}, although no standardized test set exists for such type of data. An effective strategy to deal with variability was to incorporate additional, longer-term knowledge explicitly into DNN models: \cite{7,8,9,yajie:taslp2015} study the incorporation of speaker-level i-vectors to balance the effect of speaker variability. Time Delay Neural Networks~\cite{tdnn,vijay} use wide temporal input windows to improve robustness. \cite{vesel2016sequence} extracts
long-term averages from the audio signal to adapt a DNN acoustic model. Similarly, in \cite{yajie-robust:is2015}, we learn a DNN-based extractor to model the speaker-microphone distance information dynamically on the frame level. Then distance-aware DNNs are built by appending these descriptors to the DNN inputs. We also try certain speaker and visual adaptive training strategies that result in about 20\% WER on the How-To corpus \cite{miao:is2016,gupta2017visual}. We follow the trend of including additional, longer-term knowledge but with end-to-end models rather than DNNs.

It is an important distinction that our work \textsl{does not require} localization of lip regions and/or extraction of frame-synchronous visual features (lip contours, mouth shape, landmarks, etc.), as is the case in ``traditional'' audio-visual ASR~\cite{13,17,20,19}, which has been developed mostly with a focus on noise robustness. For the majority of our data, lip-related information is not available at all, or the quality is extremely poor. Instead, we will use semantic visual features such as objects~\cite{25} and scenes~\cite{29}.

Unlike Hidden Markov Models, end-to-end systems are directly optimized for a sequence of characters, phones, or other target units.
The CTC~\cite{graves2006connectionist} loss function is defined over a target label sequence, and introduces an additional ``blank'' label, which the network can predict at any frame without influencing the output label sequence.
In CTC training, the sequence of labels are monotonically mapped to the observations (\textit{i.e.}, speech frames), and outputs appear time-aligned to inputs. Decoding of CTC models can be achieved with weighted Finite State Transducers (WFSTs,~\cite{sak2015learning,yajie-eesen:asru2015}) or a Recurrent Neural Network language model~\cite{bengio2003}. In the latter case, characters are frequently used as target labels~\cite{hannun2014first}.
%
Finally, attention-based S2S models~\cite{sutskever2014sequence} can be applied to speech recognition as well~\cite{bahdanau-s2s,chan2015listen}, and present an intriguing alternative to conventional models. Still, their training and decoding are not well understood at this point, in particular on data-sets for which high-quality annotations may not be available. We present an analysis of results with these models to better understand them.

\section{Data and Features}
\label{sec:data}

We conduct our experiments on two data-sets, the Wall Street Journal (WSJ, SI-284, LDC93S6B and LDC94S13B), and the How-To audio-visual dataset. 
The How-To corpus consists of English language open-domain instructional videos that explain specific tasks like baking a cake, or nutrition habits, and have been recorded in various environments indoors and outdoors (like kitchen, or garden), usually with a portable video recorder~\cite{miao:is2016,miao2014improvements}. Ground truth transcriptions of these videos have been created by re-aligning provided sub-titles, which sometimes are mis-matched because of missed phrases, word repetitions, hesitations, and other noise and punctuation that hasn't been transcribed.

We use 90 and 480 hours of How-To data and 87 hours of WSJ, and extract 40-dimensional MEL filter banks, with a step size of 30\,ms, 3-fold oversampling of the data at 0, 10, and 20\,ms offsets, and stacking 3 neighboring frames together, to give the same 120-dimensional input vector to both CTC and S2S models. The 90\,h subset of How-To has been selected randomly. We have a separate 4\,h test set. 5\% of the training data is used as dev set. For WSJ, we use the eval92 test set.
Both models are character-based with 43 labels/tokens: 26 alphabets, 10 digits, and special symbols for \{`.', `'', `-', `/'\}, space, start and end of sentence.

\begin{figure}[t]
  \centering
  \includegraphics[width=0.9\linewidth]{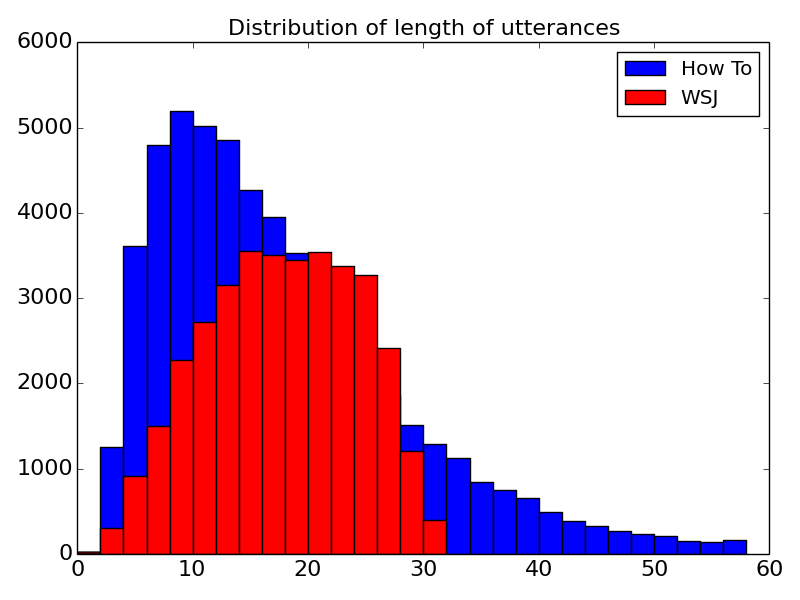}
  \caption{Length Distribution for How-To and WSJ Train set}
  \label{fig:data_distribution}
\end{figure}

Figure \ref{fig:data_distribution} shows the length distributions for How-To and WSJ train sets. This shows that for ``open-domain'' speech data, the distribution is less normalized when compared with prepared datasets. 

\subsection{Extraction of Visual Vectors}
\label{subsec:vis_features}

The visual features used in this paper are the same as those in our previous work \cite{gupta2017visual}. We extract object and place/scene features from pre-trained CNNs and perform dimensionality reduction to obtain 100 dimension features. As described above, the data contain indoor and outdoor recordings of instructional videos where object and place features are most relevant. We use these features to infer acoustic and language information from the scene where the utterance has been recorded.


\section{Models}
\label{sec:models}

    

\begin{figure*}[h]
 \begin{minipage}{\linewidth}
    \centering
    \subfloat[CTC model architecture with adaptation]{%
      \includegraphics[trim=150 0 -10 30,clip,width=0.46\linewidth,keepaspectratio]{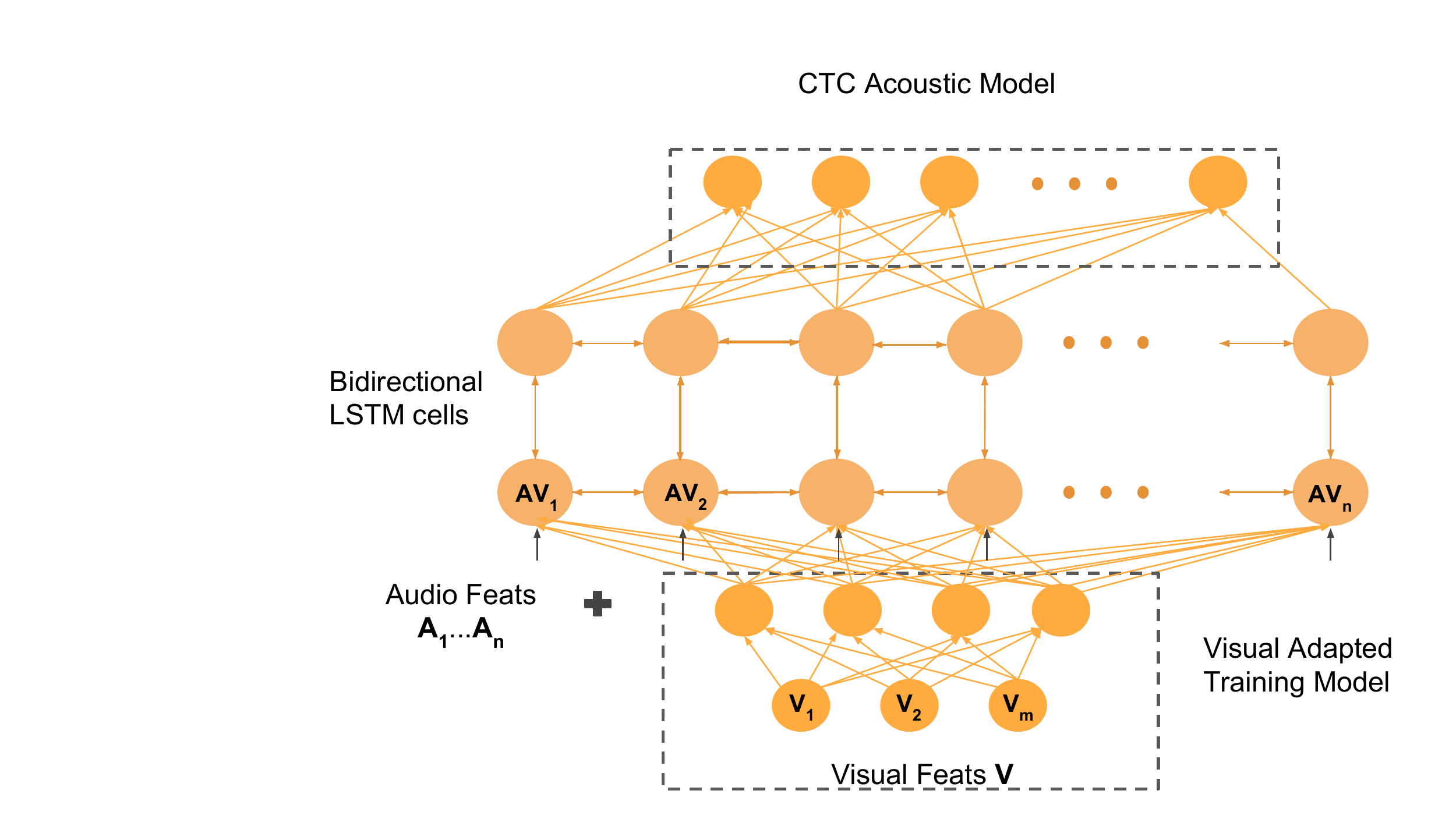}%
      \label{fig:ctc_arch}%
    }\quad%
    \subfloat[S2S model architecture with global attention and adaptation]{%
          \includegraphics[trim={0 0 100 35},clip,width=0.52\linewidth,keepaspectratio]{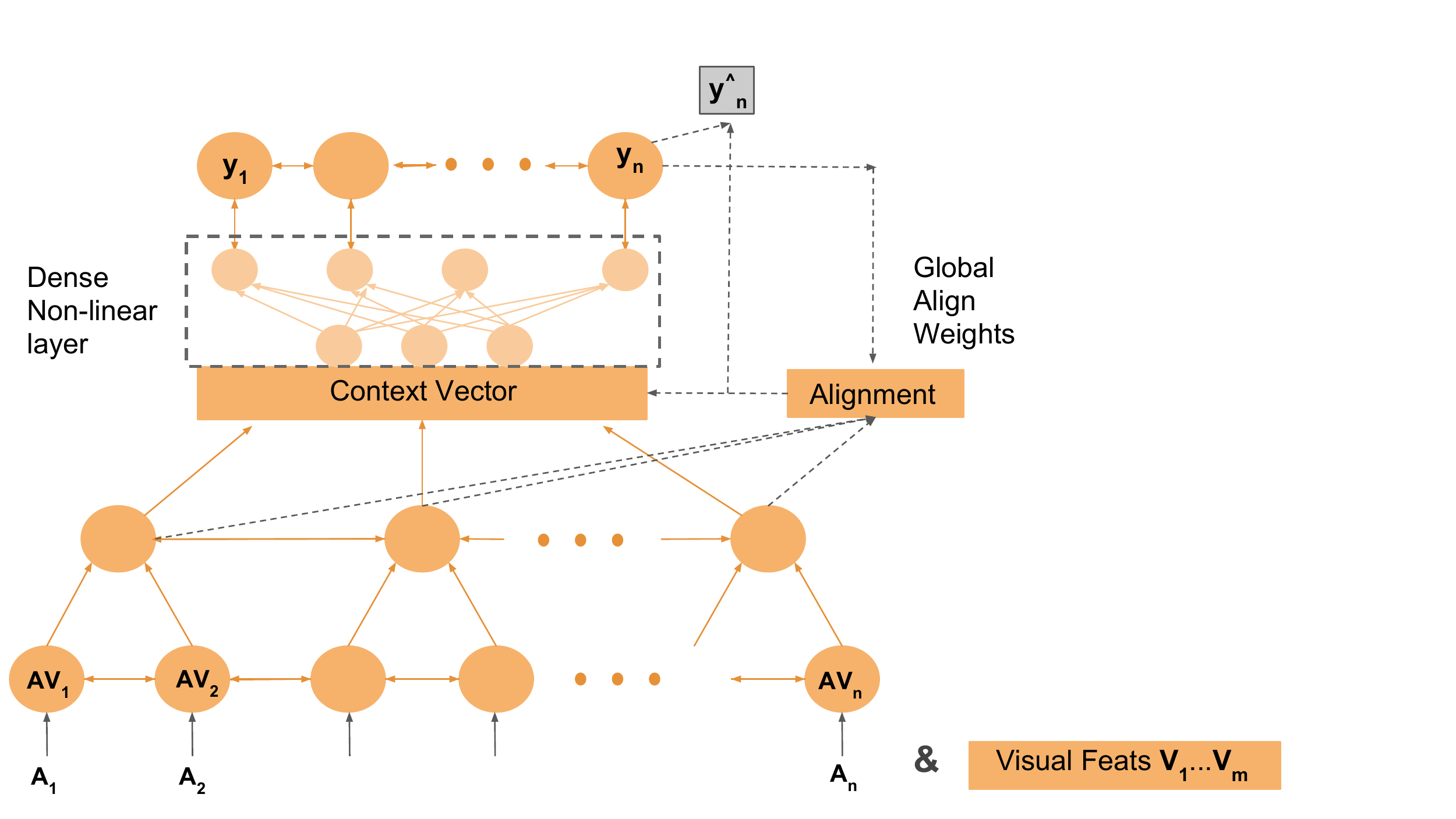}%
      \label{fig:s2s_arch}%
    }
    \caption{Model Architectures}
  \end{minipage}
\end{figure*}


\subsection{Audio-Visual ASR using CTC}
\label{ssec:avctc}


The CTC acoustic model (AM) is a stacked bidirectional Long Short Term Memory (LSTM) model with a soft-max layer at the end to generate probability of label $k \in L'$ at a particular time step $t$ given the speech sequence $\bf{X}$. Here, $L' = L \cup \phi$ where L represents the vocabulary described in Section \ref{sec:data} and $\phi$ denotes the special blank symbol introduced by CTC. To define the CTC loss function we need a many-to-one mapping $\mathcal{B}$ that maps path $\textbf{p}=(p_1,...,p_T) \in L'^T$ of the CTC model over length T to an output sequence $\textbf{z}$. The probability of $\textbf{z}$ is calculated by summing over all possible paths in CTC.

CTC AM is a frame-based sequence criterion where for every frame, a conditionally independent probability distribution over all labels $L'$ is generated. From this definition, we can infer that CTC is more dependent on the actual acoustics of the input, and can function even with poorly transcribed data. We support this hypothesis with analysis in Section \ref{sec:exp}. 

Decoding strategies to obtain labels from the probability distributions are: greedy search, WFST and character-Recurrent Neural Network (char-RNN). Greedy decoding outputs the character with the highest probability value without taking into account any previous information. We note that this strategy produces good results with the CTC AM and with visual adaptation as well. The second strategy for decoding is to re-score the output of the AM with a word-based WFST LM like our previous work \cite{yajie-eesen:asru2015}.  Char-RNN is a neural language model like in \cite{bengio2003} and provides an all neural model that can be trained end-to-end.

We implement Visual Adaptive Training (VAT) based on our previous work \cite{yajie-robust:is2015,gupta2017visual}. We jointly train a CTC AM and a separate multi-layer perceptron that is connected to the input LSTM units by a `sum' operation. This is shown in Figure \ref{fig:ctc_arch}. The intuition behind this architecture is that the VAT module will perform a normalization over the input features (audio) based on the visual information. With this adaptation, a `weighted sum' of audio and video features gives us a 120\,d features for each frame. Other context information like i-vectors, can be used with this technique. Although in our previous work we trained both models in ``two-step'' fashion (first AM, then VAT), we experienced higher boost of performance by training both models jointly, end-to-end. Other types of adaptation such as concatenation were tested but none of them outperformed the method discussed.

Our CTC AM uses 5 layers with 200 LSTM cells each, with a projection layer of 100 cells between each LSTM layer. We tried many different architectures to find this best configuration. We use Stochastic Gradient Descent (SGD) for training. Training on 480\,h How-To data takes 30 hours and greedy decoding takes less than 15 minutes on TitanX GPU with 11GB RAM.

\subsection{Audio-Visual ASR using S2S}
\label{ssec:avs2s}


We implement an attention-based S2S model that is similar to \cite{chorowski2014end,chan2015listen}. The encoder transforms the input to a high level, low dimension encoding, $h = (h_1,...,h_T)$ of length $T$, and the attention decoder produces a character probability distribution conditioned on all previous output. For each layer in the encoder, we concatenate two consecutive nodes to get a reduction factor of 2 at each layer and form a pyramidal encoder instead of a stacked one. This encoder acts as the `AM' in CTC. The attention mechanism we implement learns a weighted global context vector $W_{\alpha}$ calculated using the source hidden state $h_s$ and the \textit{current} target $h_t$ \cite{ManningAttention}. This context vector is global as it always attends to all source states $s'$. We compute a variable-length alignment vector $\alpha_t$ using: $exp(h_t^T . W_\alpha . h_s)$ and this is normalized over all input $s'$ as $\sum_{s'} exp(h_t^T . W_\alpha . h_{s'})$. The model architecture with attention is shown in the Figure \ref{fig:s2s_arch}.

We note that this type of encoder state concatenation and attention mechanism are the novelty of the S2S model in this paper and have not been applied for ASR before. The concatenation method worked slightly better in our experiments than other reduction methods reported in \cite{bahdanau-s2s,chan2015listen,KimJointCTCAttn}. In addition, a dense-nonlinear layer is used in between the encoder and decoder states. The S2S decoder implicitly learns an LM using the transcripts although this is a weak LM. Decoding is performed using beam search.

S2S is a sequence learning criterion that does not try to output a token for each frame of input. It ``attends'' over the input and output sequences to align them (multiple frames can be mapped to one token). This leads us to the hypothesis that S2S is much more dependent on the transcripts rather than the acoustics. Again, we support this hypothesis with analysis in Section \ref{sec:exp}.

In our S2S model, we use 3 layers of 512 bidirectional LSTM cells in the encoder. We use SGD with learning rate of 0.2 and decay of 0.9. We use curriculum learning \cite{bengio2009curriculum} for the first epoch to speed up convergence. We note that our training process is much simpler than \cite{chan2015listen,ChorowskiDecoding,bahdanau-s2s}. 
The decoder is made of 2 layers of 512 bidirectional LSTM cells each. For decoding, we use a beam size of 5. We do not use any techniques for better decoding with WSJ as given in \cite{ChorowskiDecoding} but use a length normalization with How-To data, to address the length distributions variance shown in Figure \ref{fig:data_distribution}. Training took 4 days with a TitanX GPU on the 90\,h subset. 
Experiments were performed using the OpenNMT toolkit \cite{opennmt}.

The video adaptation technique we use with S2S is early fusion where 100\,d visual features are concatenated with 120\,d audio features giving 220\,d vector for each frame. Our experiments with early fusion show that S2S benefits with this technique while CTC or DNN \cite{yajie-robust:is2015} does not, moreover gives better gains that VAT for CTC.

\section{Experiments and Analysis}
\label{sec:exp}

\subsection{Visual Feature Adaptation}
\label{ssec:adaptation}
In Table \ref{tab:adaptation_results}, we present the effect of adaptation on the Token Error Rate (TER) and Perplexity (PPL) of the two models. We show that the adaptation with visual features helps improve the absolute TERs by 1\% in CTC and by 1.6\% in the S2S model. Using length norm as in Section \ref{ssec:wsj_ht}, we improve the S2S improvement to 2\%. In our experience, this is a significant improvement in such sequence-based models. The PPL values for CTC are those of a word LM while those of S2S are of the implicit character LM (*). We see that adapting a language model with visual features helps decrease the perplexity by a huge margin. This establishes that there is a strong correlation between visual features and speech. 
The perplexity for S2S models is the character-prediction perplexity (joint AM, LM) and we see that there is no difference in this case. We calculate TER as our experiments with greedy decoding for CTC AM showed good results with the visual adaptation, and the implicit LM of S2S was also quite strong, as we show below. The `dev' set is a tougher set than the `test' set in the How-To dataset.
\begin{table}[h!]
\centering
\begin{tabular}{ c c c c c}
\hline
			& \textbf{A} CTC & \textbf{A+V} CTC & \textbf{A} S2S & \textbf{A+V} S2S \\ \hline
TER dev 	& 	15.2 & 	14.1	& 18.4		& 16.8	\\
TER test 	& 	13.6 & 	13.1 & 16.3		& 15.7	\\
PPL* dev 	& 	113.6  & 80.6 & 1.38		& 1.37  \\
PPL* test 	& 	112.0	& 	72.0 & 1.05		& 1.05  \\
\hline
\end{tabular}
\caption{Results for Audio(\textbf{A}) and Audio-Visual(\textbf{A+V}) adaptation with the How-To data}
\label{tab:adaptation_results}
\end{table}

\subsection{CTC vs. S2S}
\label{ssec:ctc_s2s}
We compare the audio-only CTC and S2S models trained on WSJ and 90\,h subset of How-To in Table \ref{tab:ctc_s2s}. TER with CTC and S2S on WSJ is a strong baseline compared to prior work \cite{hannun2014first,graves-jaitly-ctc, bahdanau-s2s,KimJointCTCAttn}. We see a huge disparity in ASR for clean prepared data (WSJ) and real application data (How-To) as discussed in Section \ref{sec:data}. We also note that for noisy data like How-To, CTC required 5 times as much data to make the error rates between S2S and CTC comparable.

\begin{table}[h!]
\centering
\begin{tabular}{ c c c } 
\hline
&  CTC & S2S \\ 
\hline
WSJ & 6.9 &  7.9  \\
How-To & 18.5 & 15.3  \\
\hline
\end{tabular}
\caption{TER of CTC and S2S on WSJ (rval92), How-To(test set)}
\label{tab:ctc_s2s}
\end{table}

Table~\ref{tab:test_eg} shows an example from the How-To test set which highlights the fundamental difference between the CTC and S2S model when applied to ``real'' data: while the TER for both models is similar in value, it stems from different types of mistakes. The CTC model's output is closer to the acoustics and produces characters even for corrections and false starts (even though those were also not present in the training transcripts, according to our analysis); there is a need for language model re-scoring. The S2S model however takes more liberty in character prediction and the resulting output is a combination of the acoustics, and the language model, which has been trained to ignore false starts. Even on an S2S system that has been trained on 90\,h only, it appears that the inbuilt language model is strong and does not produce out of vocabulary words like CTC does, similar to the effect we observed when decoding CTC AMs with an RNN LM on the Switchboard corpus~\cite{is2017:ctc}.

\begin{table}[h!]
\centering
\begin{tabular}{ p{1.5cm} p{6.5cm} }  
\hline
Spoken & now it does only say for do- or doesn't even say for dogs or cats it's neither
  \\ \hline
Reference & now \textbf{doesn't even} say dogs or cats \textbf{it says} neither
 \\ \hline
Greedy CTC & now it does only say\textbf{e} for do\textbf{g or toat use a} dogs or cats \textbf{s niter} \\ \hline
S2S & now it does\textbf{n't we} say for \textbf{a} do\textbf{g or that use a} dogs or cats \textbf{so is night or}  \\ \hline
AV S2S & now it does\textbf{n't leave safer} do\textbf{g or it does use a} dogs or cat \textbf{so in night or} \\
\hline
\end{tabular}
\caption{Typical transcription on How-To test set: the CTC model is close to acoustics even
during a correction (following ``do-''), while S2S keeps to the style of the reference, which is itself an abstraction of the spoken content. Currently, there is little semantic difference between regular and adapted (AV) S2S (or CTC) output.}
\label{tab:test_eg}
\end{table}

\subsection{WSJ vs. How-To in S2S}
\label{ssec:wsj_ht}
\begin{figure}
\centering
\includegraphics[width=7cm]{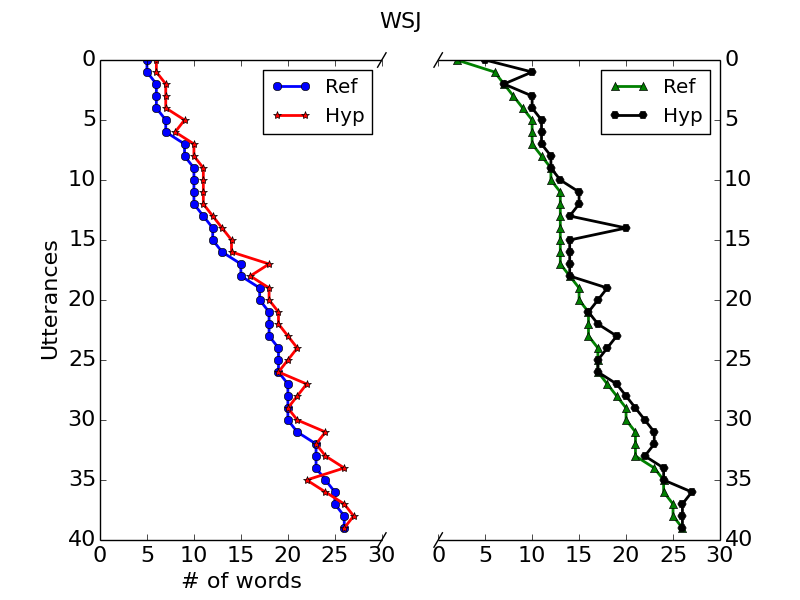}
\includegraphics[width=7cm]{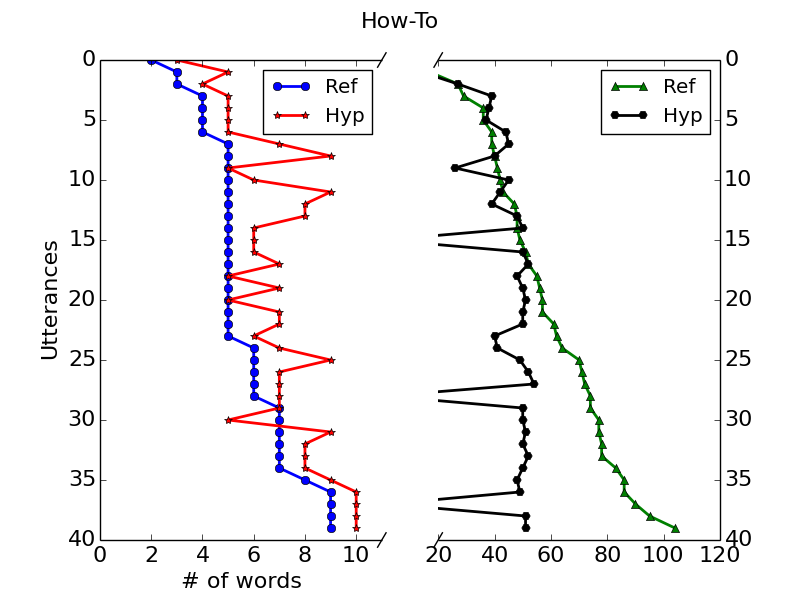}
\caption{Length normalization by S2S for WSJ and How-To}
\label{fig:length_norm}
\end{figure}

Figure~\ref{fig:length_norm} compares reference length to hypothesis length for 40 short and long WSJ and How-To test utterances.
On WSJ, the range of lengths of short and long utterances are similar, and reference and hypothesis follow each other closely. On How-To, hypothesis prediction is very unstable and the model makes a lot of mistakes, even breaks completely at times. Length of the hypothesis is greater than the length of reference for short utterances, while it is lesser for longer utterances. As seen from the example in Table \ref{tab:test_eg}, the output of the S2S model is much closer to the reference transcript. The model learns a form of length normalization over the entire dataset hence performs badly on short and long utterances. We use the length normalization factor during decoding to stabilize the output of the S2S model and get absolute improvements of 2\% (dev) and 1\% (test) for the non-adapted case, which is slightly better than the adapted case and shows that adaptation stabilizes model performance.  With our model we see no need for normalization on WSJ, different from \cite{ChorowskiDecoding}.

\section{Conclusions and Future Work}

In this paper, we describe and compare two end-to-end models that learn semantically relevant context information from the visual channel of a video, and use it to improve speech transcription. We show how to adapt a CTC bidirectional LSTM acoustic model and a S2S model to the visual semantic features. We compare the behavior of CTC and S2S models on a clean (WSJ) and noisy (How-To) dataset, and see that CTC output tends to be very close to the acoustics of an utterance, while S2S output appears to be closer to the style of the transcriptions. Similar to \cite{corr/abs-1711-07274}, we find that S2S approaches are maybe surprisingly robust against ``real-world'' data, that has not been carefully prepared for speech recognition experiments. On the WSJ dataset, our system outperforms previous S2S implementations like \cite{KimJointCTCAttn}. Next, we would try more adaptation strategies like the MLP shift with S2S model and re-scoring output of CTC with a char-RNN LM to form an all neural model.

\section{Acknowledgments}
\ninept
We gratefully acknowledge the support of NVIDIA Corporation with the donation of the Titan X Pascal GPU used for this research.
This work used the Extreme Science and Engineering Discovery Environment (XSEDE), which is supported by National Science Foundation grant number OCI-1053575.  Specifically, it used the Bridges system, which is supported by NSF award number ACI-1445606, at the Pittsburgh Supercomputing Center (PSC).

\bibliographystyle{IEEEbib}
\bibliography{strings,refs}

\end{document}